\documentclass[article,moreauthors,pdftex,10pt,a4paper]{mdpi} 
\firstpage{1} 
\pdfoutput=1
\Title{L\'evy analysis of HBT correlation functions in $\sqrt{s_{\rm NN}}=$ 62 GeV and 39 GeV Au+Au collisions at PHENIX}

\Author{Dániel Kincses\orcidA{} for the PHENIX Collaboration}
\AuthorNames{Dániel Kincses}
\address[1]{Eötvös Loránd University, H-1117 Budapest, Pázmány P. s. 1/A, Hungary}
\corres{Correspondence: kincses@ttk.elte.hu; Tel.: +36-1-372-2500/6037}

\abstract{The phase diagram of strongly interacting matter can be explored by analyzing data of heavy-ion collisions at different center of mass collision energies. For investigating the space-time structure of the hadron emission source, HBT correlation measurements are among the best tools. In this paper we present the latest results from the RHIC PHENIX experiment on such measurements in $\sqrt{s_{\rm NN}}=$ 39 GeV and 62 GeV Au+Au collisions.}

\keyword{RHIC, PHENIX, femtoscopy, HBT, Bose-Einstein correlations, Lévy distribution, anomalous diffusion, phase diagram, critical point}

\conferencetitle{10th Bolyai-Gauss-Lobachevsky Conference on Non-Euclidean Geometry and its Applications}

\usepackage{subcaption}
\usepackage{chngcntr}
\preto{\abstractkeywords}{\nolinenumbers}
\begin{document}

\section{Introduction}

Experimental data from heavy-ion collisions are in agreement with the theoretical prediction~\citep{Aoki:2006we} that the quark-hadron transition is cross-over at high energies (near zero baryochemical potential). At lower energies the transition is expected to be a first-order phase transition. This indicates that there could be a critical endpoint somewhere between, where second-order phase transition takes place. Finding and characterizing this critical point is one of the most important current problems in heavy-ion physics. To study the phase diagram of strongly interacting matter, we have to perform measurements at different collision energies so that we can gain information about different regions on the phase diagram. In the past years the PHENIX experiment at the Relativistic Heavy Ion Collider of the Brookhaven National Laboratory collected a vast amount of data at various collision energies. The gold-gold beam energy scan spans from 7.7 GeV up to 200 GeV in $\sqrt{s_{NN}}$. In terms of baryochemical potential the range is going up to approximately 400 MeV, while in freeze-out temperature this range is between 170 MeV and 140 MeV~\citep{Adamczyk:2017iwn}. An illustration of the phase-diagram can be seen on Figure \ref{ff:QCDphases}, while the recently extracted chemical freezeout parameters are shown on Figure \ref{ff:chemfr}.

There is a very practical tool to gain information about the particle-emitting source, namely the measurement of Bose-Einstein or HBT correlations of identical bosons \citep{Wiedemann:1999qn,Csorgo:1999sj,Lisa:2005dd}. In the previous years usually Gaussian-type sources were utilized to describe the correlation functions, but the latest results from the PHENIX experiment showed that we can and have to go beyond the Gaussian approximation~\citep{Csanad:2005nr,Adare:2017vig}. In an expanding medium, on the basis of the generalized central limit theorem and the anomalous diffusion we can expect the appearance of L\'evy-type sources~\citep{Csanad:2007fr,Csorgo:2003uv,Metzler:1999zz}. The symmetric L\'evy distribution is the generalization of the Gaussian distribution and it is defined by the following expression: 

\begin{equation}
\displaystyle \mathcal{L}(\alpha,R,r)=\frac{1}{(2\pi)^3} \int d^3q e^{iqr} e^{-\frac{1}{2}|qR|^{\alpha}}.
\end{equation}

The $R$ parameter is called the L\'evy scale parameter, and the $\alpha$ parameter is the index of stability or L\'evy exponent. In case of $\alpha=2$ we get back the Gaussian distribution, while the $\alpha=1$ case results in the Cauchy distribution. In case of $\alpha < 2$ the L\'evy distribution exhibits a power-law behavior. Using plane-wave approximation (neglecting the final-state interactions) and assuming that the source is a spherically symmetric three-dimensional L\'evy distribution, furthermore using the framework of the core-halo model, the two-particle correlation function takes the following simple form ($Q$ is a one-dimensional relative momentum variable, see details in~\citep{Csorgo:2003uv}):

\begin{equation}
\large{C_{2}(Q)=1+\lambda\cdot e^{-(QR)^{\alpha}}}.
\label{C0}
\end{equation}

The L\'evy-type source also gives a hand in the search for the critical point, as one of its parameters, the index of stability $\alpha$ is related to one of the critical exponents, $\eta$. 
In case of a second order phase transition the $\eta$ exponent describes the power-law behavior of the spatial correlations at the critical point with an exponent of $-(d-2+\eta)$ where $d$ is the dimension. In case of the three-dimensional L\'evy distribution the power-law tail of the spatial correlations is described with an exponent of $-1-\alpha$. From comparing these exponents we can easily see that the L\'evy exponent is identical to $\eta$ at the critical point~\citep{Csorgo:2009gb}. The second-order QCD phase-transition is expected to be in the same universality class as the 3D Ising model~\cite{Halasz:1998qr,Stephanov:1998dy}, and in that case the $\eta$ exponent has a definite value of 0.03631(3) at the critical point~\citep{El-Showk:2014dwa}. Some argue~\citep{Csorgo:2005it} that QCD more likely falls in the universality class of the random field 3D Ising model, where the value of $\eta$ is 0.5$\pm$0.05 \cite{Rieger:1995aa}. If one extracts the value of the L\'evy exponent $\alpha$ at different center-of-mass collision energies, the data may yield information on the nature of the quark-hadron phase transition, particularly it may shed light on the location of the critical endpoint (CEP) on the phase-diagram. 
 \vfill
{\centering
\begin{minipage}{.495\linewidth}
	\begin{figure}[H]
    \includegraphics[height=0.85\linewidth]{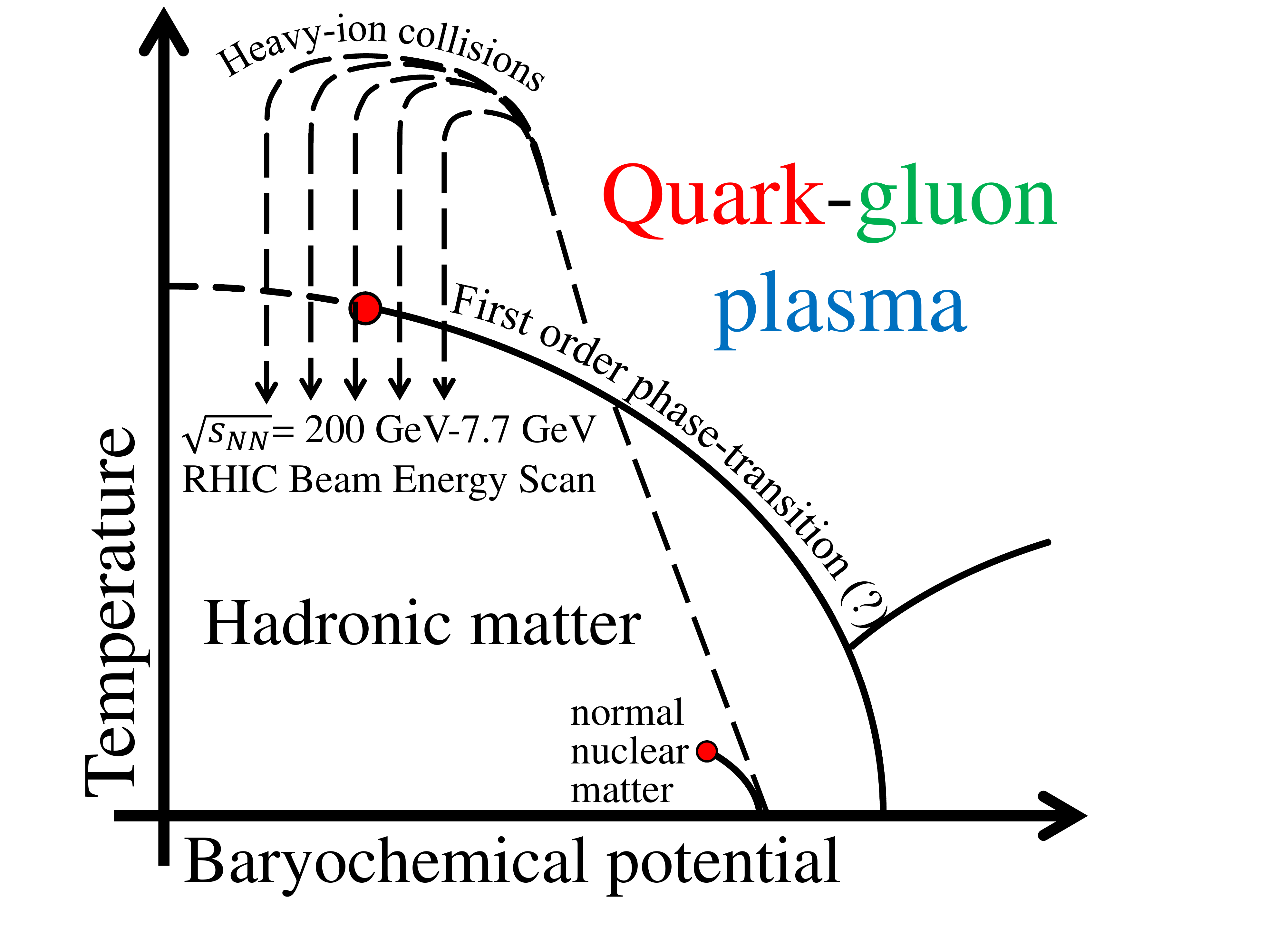}
    \caption{The phase diagram of strongly interacting matter.}
	\label{ff:QCDphases}
	\end{figure} 
\end{minipage}
\hfill
\begin{minipage}{.495\linewidth}
	\begin{figure}[H]
    \includegraphics[height=0.85\linewidth]{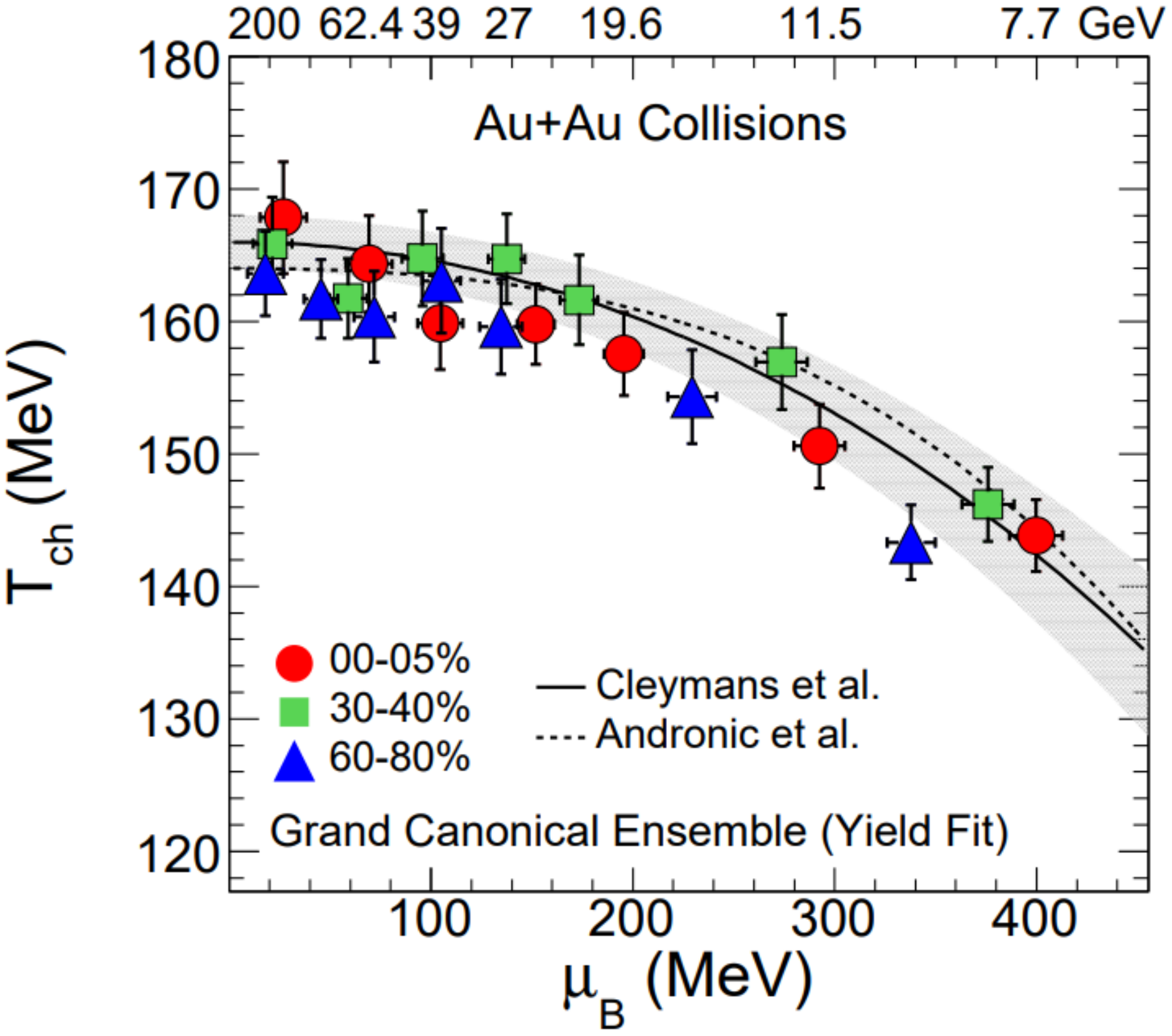}
    \caption{Extracted chemical freezeout parameters from Ref. \citep{Adamczyk:2017iwn}}
    \label{ff:chemfr}
    \end{figure} 
\end{minipage}}
\vfill
\newpage
\section{Results}
	
In our analysis we used $\sqrt{s_{NN}}=62$ GeV and $\sqrt{s_{NN}}=39$ GeV Au+Au data recorded by PHENIX during the 2010 running period. We measured two-particle HBT correlation functions for $\pi^+\pi^+$ and $\pi^-\pi^-$ pairs. In case of the 62 GeV dataset we used 8 average transverse mass ($m_T$) bins and 4 centrality ranges, while in case of the 39 GeV dataset we used 6 $m_T$ bins and 2 centrality ranges. We checked that the correlation functions of $\pi^+\pi^+$ and $\pi^-\pi^-$ pairs are consistent, so we combined the data. The aforementioned formula for the correlation function (Eq. \ref{C0}.) does not take into account the final-state Coulomb interaction. In our fits we incorporated that effect and also introduced a linear background. In all $m_T$ bins the fits described the measured correlation functions in a statistically acceptable manner. 

\begin{figure}[H]
	\centerline{
	\includegraphics[width=0.49\textwidth]{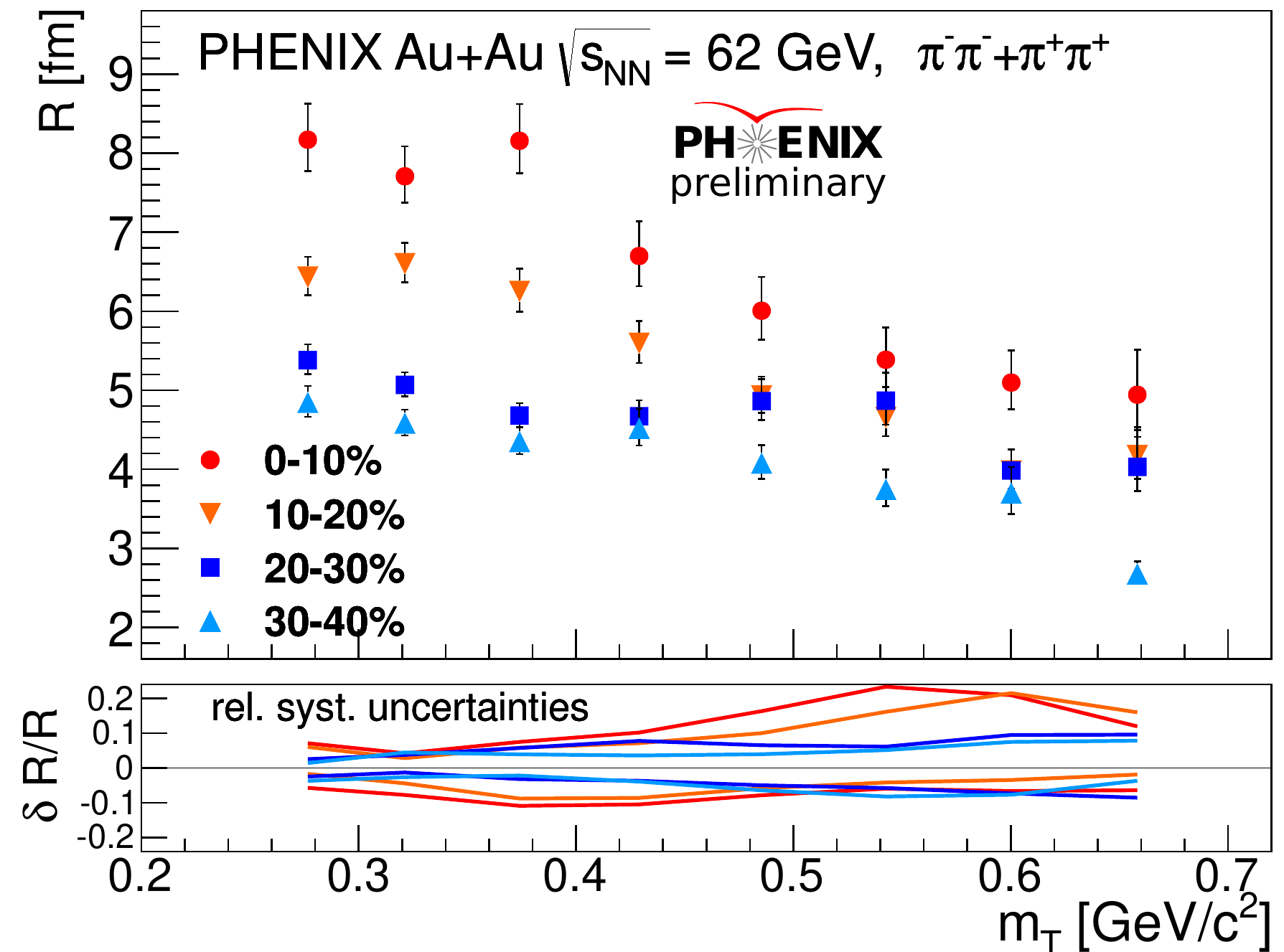}
	\includegraphics[width=0.49\textwidth]{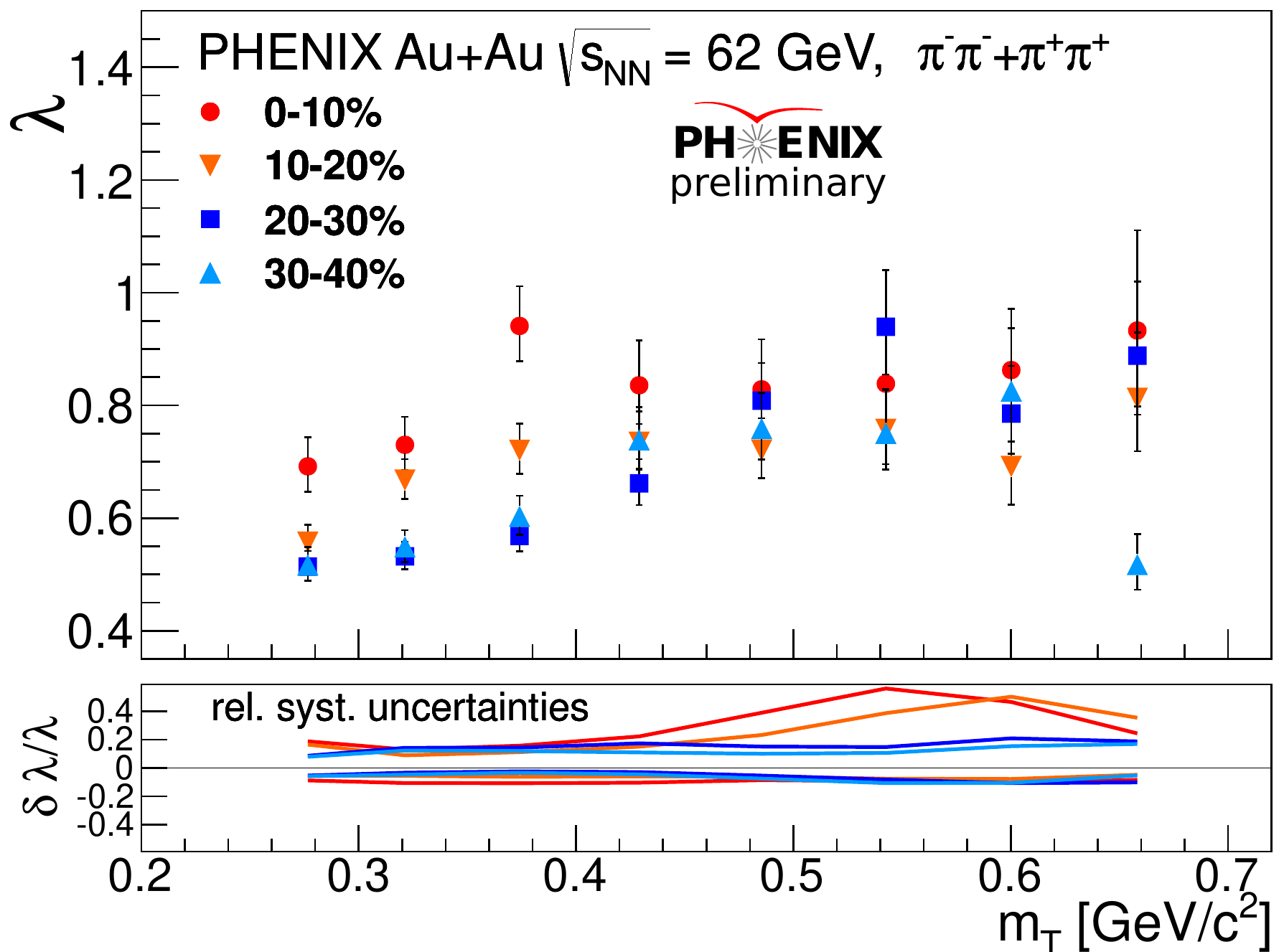}}
	\centerline{
	\includegraphics[width=0.49\textwidth]{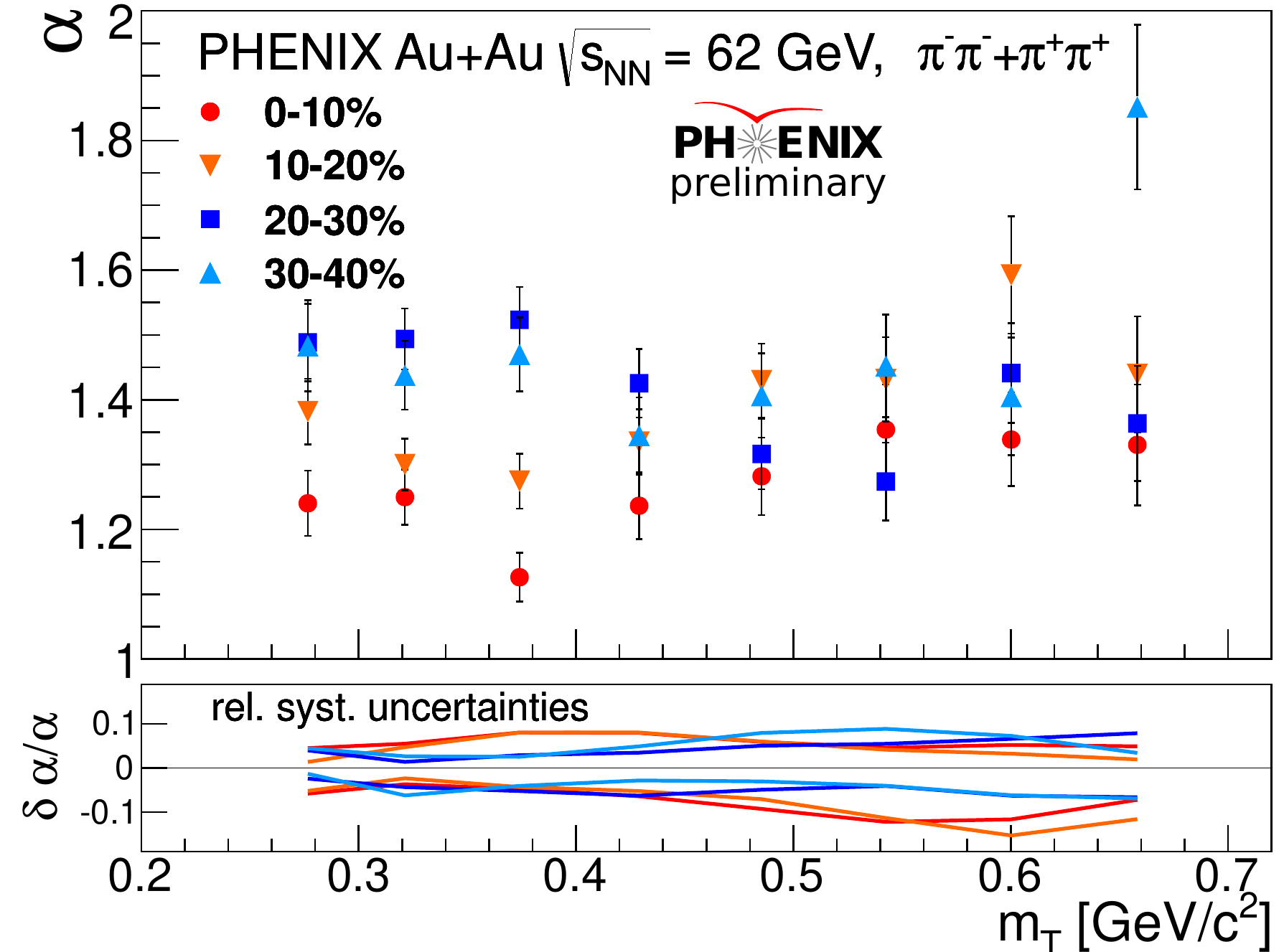}
	\includegraphics[width=0.49\textwidth]{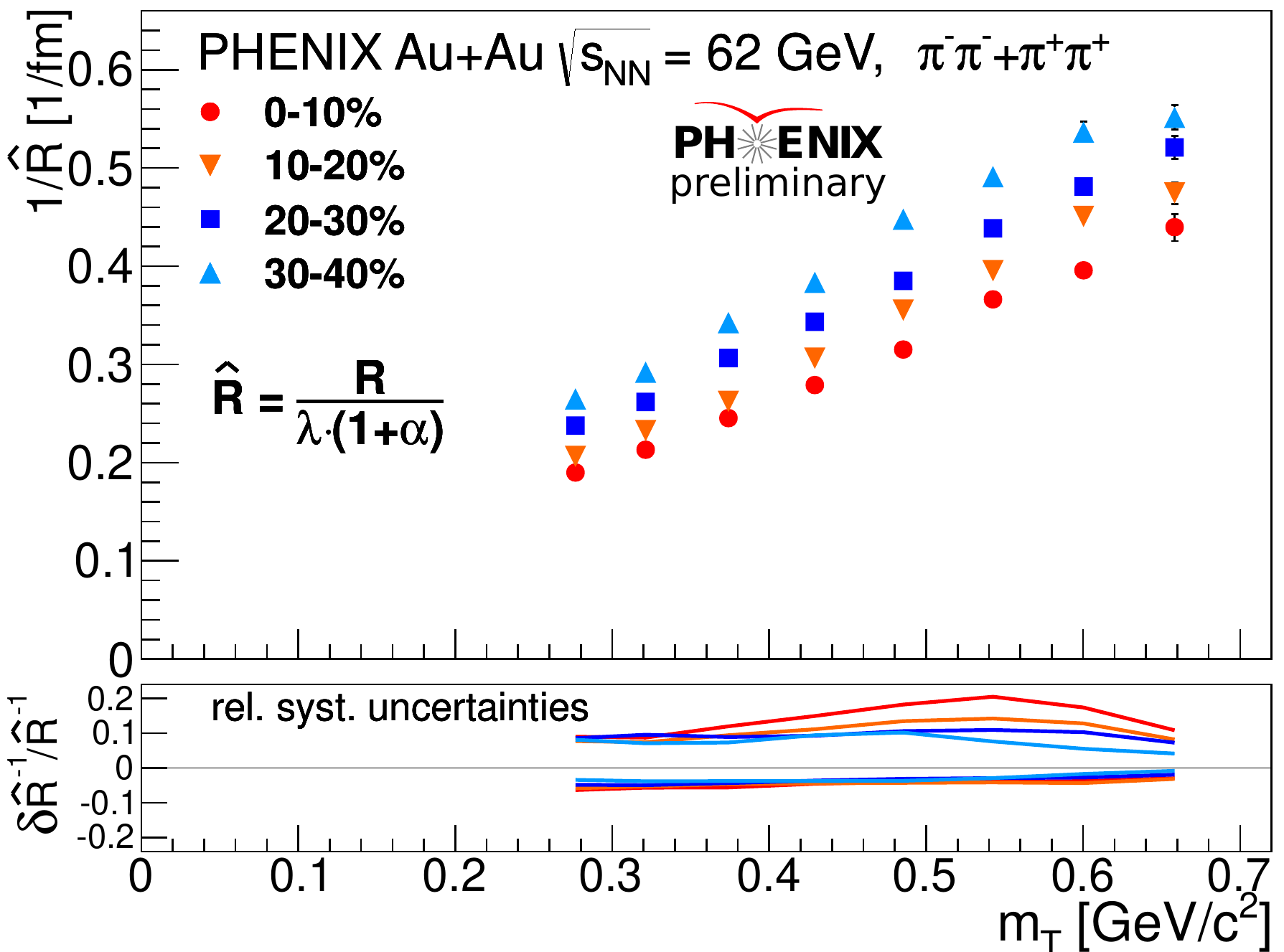}}
	\centerline{
	\includegraphics[width=0.49\textwidth]{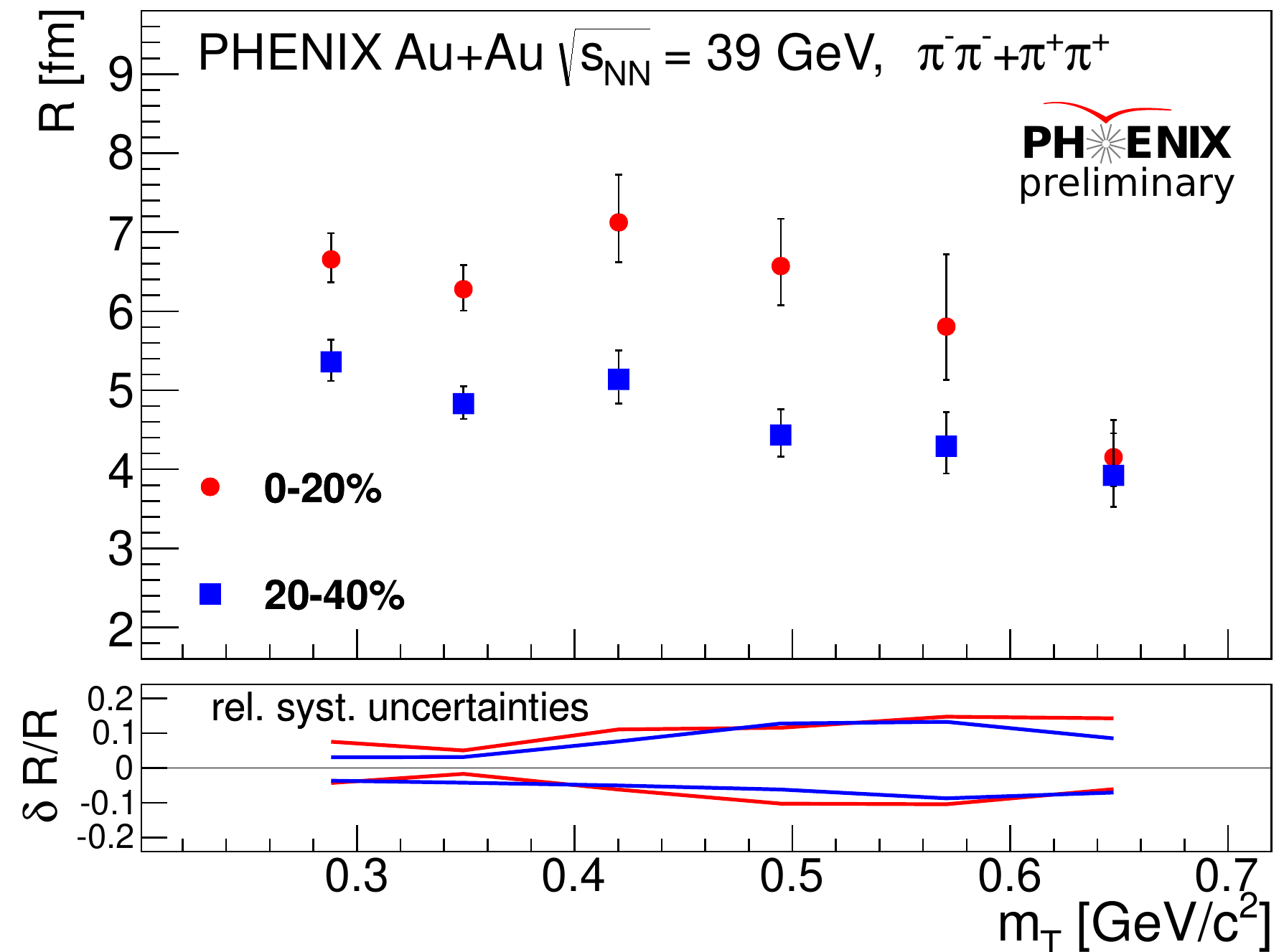}
	\includegraphics[width=0.49\textwidth]{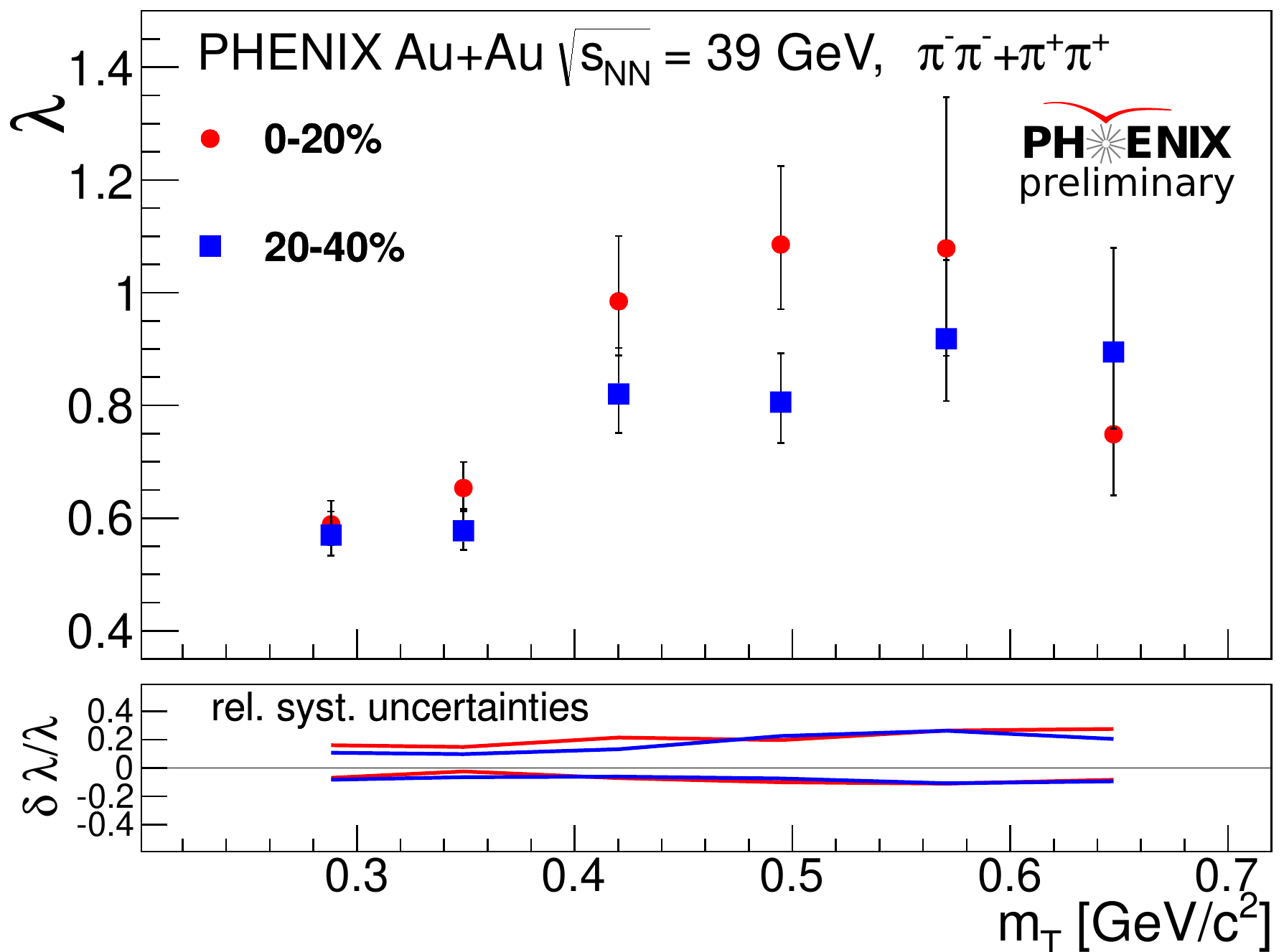}}
	\centerline{
	\includegraphics[width=0.49\textwidth]{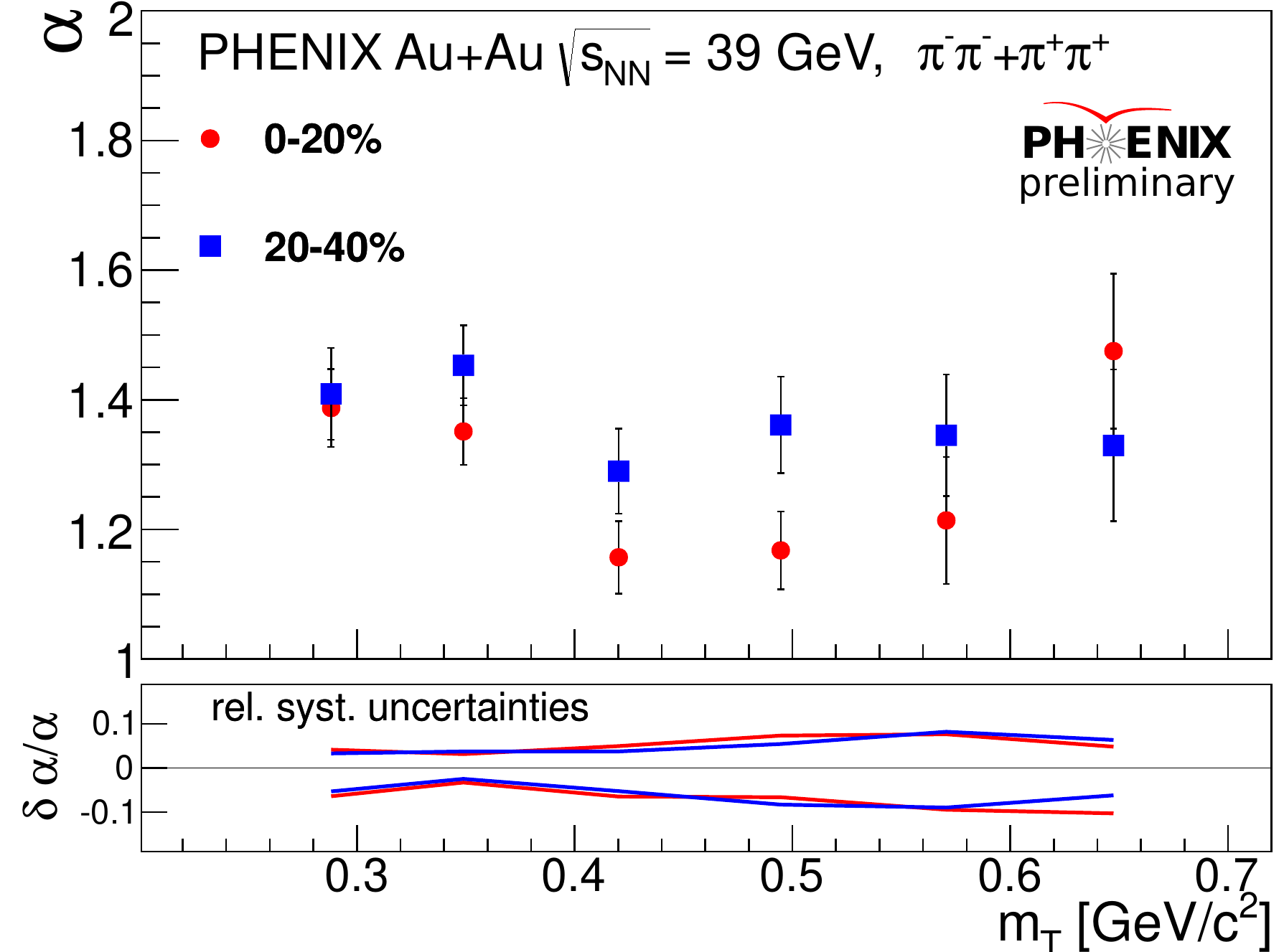}
	\includegraphics[width=0.49\textwidth]{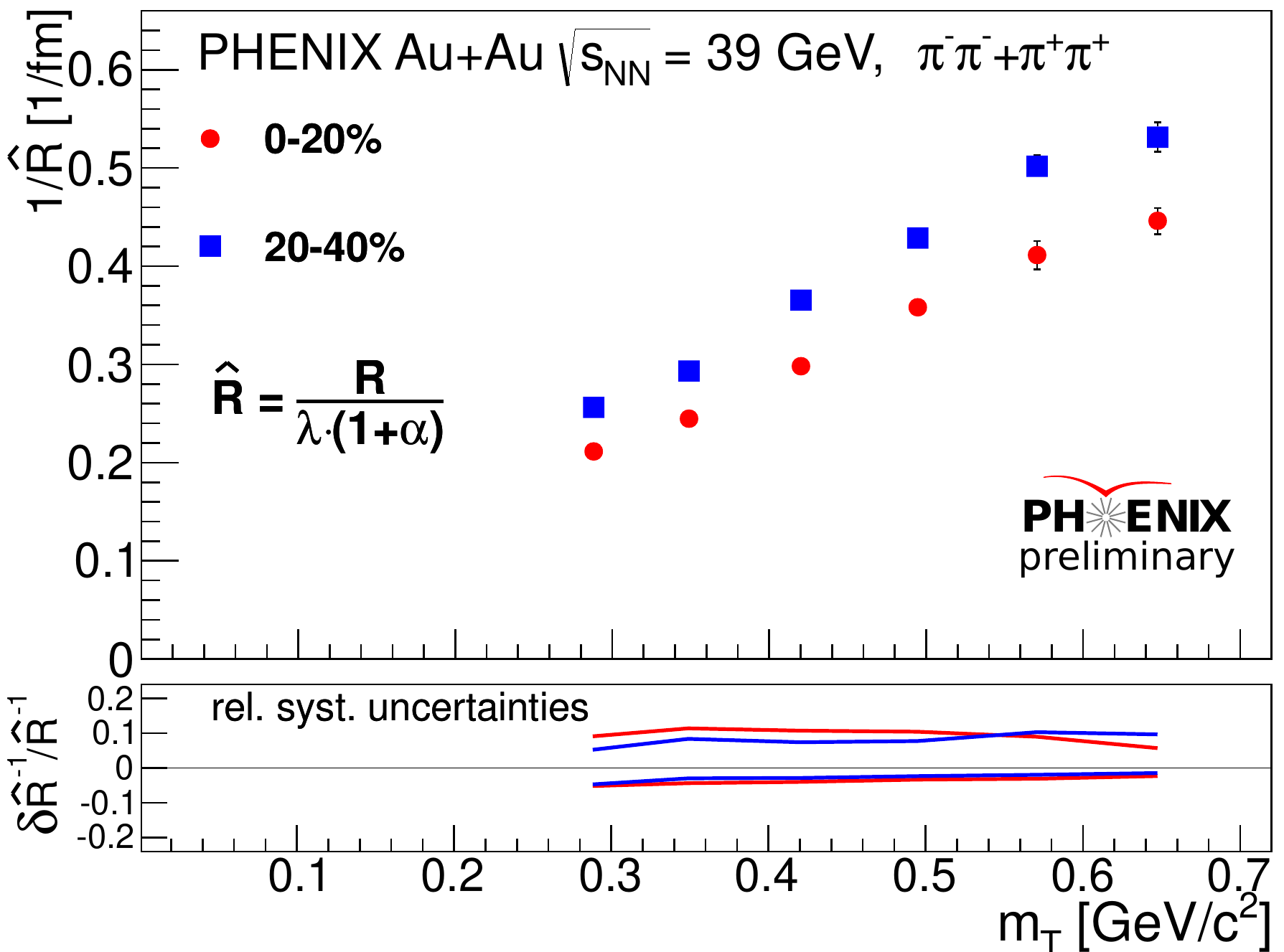}}
	\caption{Centrality and $m_T$ dependence of L\'evy source parameters in $\sqrt{s_{NN}}$ = 62 GeV and $\sqrt{s_{NN}}$ = 39 GeV Au+Au collisions. The different colors and marker styles are denoting the different centrality ranges. The auxiliary figures at the bottom show the relative systematic errors.}
	\label{ff:results}
	\end{figure}

In Figure \ref{ff:results} we present the resulting fit parameters ($R$,$\lambda$,$\alpha$) versus pair $m_T$, as well as an empirically found scaling parameter $\widehat{R}$. We observe that most features of the 200 GeV results~\citep{Csanad:2017usp} are resembled by these low energy data. The L\'evy scale parameter shows a decreasing behavior, and we can clearly observe a geometrical centrality dependence. In more central collisions, the $R$ parameter (which is related to the physical size of the system) takes larger values. According to the core-halo model~\cite{Bolz:1992hc,Csorgo:1994in}, the correlation strength parameter is in connection with the core-halo ratio, $\lambda = (N_C/(N_C+N_H))^2$. In this formula $N_C$ denotes the number of pions created from the freeze-out of the quark-gluon plasma (or from the decay of very short lived resonances), and $N_H$ is the number of pions created in the decays of long-lived resonances. Just like in the 200 GeV case, we can observe a decrease in $\lambda$ values at small $m_T$ at both 62 and 39 GeV which may point to the increase of the halo fraction in that region. In case of a partial chiral $U_A(1)$ restoration, the in-medium mass of the $\eta'$ meson could be modified~\citep{Kapusta:1995ww} which then could lead to an increase in its yield. This effect would result in an increased number of halo pions in the small-$m_T$ region, and hence it could provide one explanation for our observation~\cite{Vance:1998wd}. Of course there could be other effects~\cite{Weiner:1999th,Csorgo:1999sj} that would result in similar behavior of $\lambda$ so a precise measurement is important. Moving on to the L\'evy exponent $\alpha$ we can see that for all centrality and $m_T$ ranges the values are between 1 and 2 for both 39 and 62 GeV. The measured values are far from the Gaussian $\alpha=2$ case and also far from the conjectured $\alpha\leq 0.5$ at the critical point. Although we have not seen an indication for critical behavior in neither the 62 GeV nor the 39 GeV measurements, we can not exclude these ranges from the critical point search because finite-size and finite-time effects can modify the picture immensely as discussed in Ref. \citep{Lacey:2015yxg,Lacey:2016tsw}. We also show an empirically found scaling parameter, defined as $\widehat{R} = R/(\lambda\cdot(1+\alpha))$. The inverse of this parameter shows a clear linear scaling with $m_T$ and also has much smaller systematic uncertainties than the other three parameters. The physical interpretation of this parameter is still an open question.
	
\vspace{6pt} 
\acknowledgments{This research was supported by the funding agencies listed in Ref. \citep{Adare:2017vig}. In addition, D. K. was supported by the New National Excellence program of the Hungarian Ministry of Human Capacities as well as the NKFIH grant FK-123842.}


\end{document}